\documentclass[%
prl,
superscriptaddress,
twocolumn,
bibnotes,
amsmath,amssymb,
longbibliography
]{revtex4-2}

\usepackage{footmisc}
\usepackage{graphicx}% Include figure files
\usepackage{epstopdf}
\usepackage{xcolor}
\usepackage{siunitx}

\usepackage{dcolumn}% Align table columns on decimal point
\usepackage{bm}% bold math
\usepackage{titlesec}

\usepackage{filecontents}

\newcommand{\ru }{$\alpha\text{-RuCl}_{3} \thickspace$}
\newcommand{\ruu}{$\alpha\text{-RuCl}_{3}$}
\newcommand{\LR}{$I_{+-}~$}
\newcommand{\RL}{$I_{-+}~$}

\titleformat{\section}{\raggedright\fontsize{12.5}{25}\bfseries}{\arabic{section}.}{1em}{}
\pdfoutput=1

\begin{document}

\pagenumbering{arabic}

%\title{\fontsize{15}{19}\selectfont Chiral excitations and the intermediate-field spin-liquid regime in the Kitaev magnet \ru }
%m Why play with the format? 
\title{Chiral excitations and the intermediate-field spin-liquid regime in the Kitaev magnet \ru }

\author{Anuja Sahasrabudhe}
\email{sahasrabudhe@ph2.uni-koeln.de}

\affiliation{Institute of Physics II, University of Cologne, 50937 Cologne, Germany}

\author{Mikhail A. Prosnikov}
\affiliation{High Field Magnet Laboratory (HFML-EMFL),
Radboud University, 6525 ED Nijmegen, The Netherlands}

\author{Thomas C. Koethe}

\author{Philipp Stein}

\affiliation{Institute of Physics II, University of Cologne, 50937 Cologne, Germany}

\author{Vladimir~Tsurkan}
\affiliation{University of Augsburg, 86159 Augsburg, Germany}
\affiliation{ Institute of Applied Physics, MD 2028, Chisinau, Republic of Moldova}

\author{Alois Loidl}
\affiliation{University of Augsburg, 86159 Augsburg, Germany}
\author{Markus  Gr\"uninger}

\affiliation{Institute of Physics II, University of Cologne, 50937 Cologne, Germany}

\author{Hamoon Hedayat}
\email{hedayat@ph2.uni-koeln.de}
\author{Paul H. M. van Loosdrecht}
\email{pvl@ph2.uni-koeln.de}

\affiliation{Institute of Physics II, University of Cologne, 50937 Cologne, Germany}

\date{May 4 , 2023}

\begin{abstract}
In the Kitaev magnet \ruu, the existence of a magnetic-field-induced quantum spin liquid phase and of 
anyonic excitations are discussed controversially. We address this elusive, exotic phase via helicity-dependent 
Raman scattering and analyze the Raman optical activity of excitations as a function of magnetic field and temperature. 
The hotly debated field regime between \SI{7.5}{\tesla} and \SI{10.5}{\tesla} is characterized by clear spectroscopic signatures such as a plateau of the Raman optical activity of the dominant, chiral spin-flip excitation. This provides direct evidence for the existence of a distinct intermediate field regime compatible with a 
quantum spin liquid phase.
\end{abstract}

\maketitle

More than a decade ago, Kitaev proposed an exactly solvable model for a spin-$\frac{1}{2}$ system on a honeycomb lattice with bond-directional nearest-neighbour exchange couplings that hosts a quantum spin liquid (QSL) ground state \cite{kitaev2006anyons}.
The collective excitations above this ground state are gapless Majorana fermions and visons. In the presence of an external magnetic field, the Majorana fermions are gapped. The gapped spectrum hosts a chiral Majorana mode responsible for thermal edge transport and the emergent excitations are non-abelian anyons ~\cite{kitaev2006anyons}.

Strong experimental efforts have been triggered by the possibility to achieve dominant Kitaev exchange 
	and the related exotic excitations in real materials \cite{JK2009materials}.
In this context, \ru has been touted as a potential candidate that could possess a QSL phase \cite{ HSKim2015RuCl3,Banerjee2017NeutronT, sandilands2015scattering,Nasu2016Raman,SHDo2017majorana, baek2017evidence}, which in turn has sparked a flurry of research in exploring its physical properties \cite{johnson2015zzorder, Cao2016zzorder, banerjee2018excitations,balz2019finite,modic2021scale}. Unlike an ideal QSL that remains disordered even at vanishing temperature, the presence of additional exchange interactions in \ru yields antiferromagnetic zigzag (ZZ) order
below \SI{7}{\kelvin} and for external magnetic fields below \SI{7}{\tesla} 
\cite{johnson2015zzorder, Cao2016zzorder, banerjee2018excitations, wagner2022magneto}. 
For strong enough magnetic fields a field induced spin-polarized phase is observed \cite{sahasrabudhe2020high, ponomaryov2020nature, wang2017magnetic}. It is still possible though that between these high- and low-field phases, bond-directional interactions take over in an intermediate field regime (IFR), thereby preparing a conducive ground for the appearance of a QSL. 

In this purported intermediate field regime, thermal conductivity measurements \cite{kasahara2018majorana, yokoi2021half,yamashita2020sample} have displayed a direct signature of a chiral Majorana edge mode (and 
hence of a QSL) -- viz.\ a half-integer quantized plateau in the transverse heat conductivity \cite{Joji2021FQH, Vinkler2018phonons, ye2018quantization}. 
However, half-integer quantisation of the thermal conductivity and consequently the role of Majorana fermions as heat carriers has been challenged \cite{Lefran2022phononQHE, Czajka2021oscTQHE, Bruin2022TQHE}.
Meanwhile, thermodynamic experiments like thermal expansion, magnetostriction, and the magnetocaloric effect provide conflicting conclusions about the presence of the QSL phase in the field regime of \SIrange{7.5(5)}{11}{\tesla} for field 
applied parallel to the honeycomb plane
\cite{balz2019finite, gass2020field, bachus2020thermodynamic, schonemann2020thermal}. 
Recent evidence in favor of the QSL phase comes from the observation of oscillations of the thermal conductivity in this field regime \cite{Czajka2021oscTQHE}.

When studied using spectroscopic experiments, \ru shows a low-energy multi-particle continuum proposed to be of Majorana fermions \cite{sandilands2015scattering, Nasu2016Raman, TTMai2019continuum}. 
Curiously, this continuum is already present in the magnetically ordered phase in zero field.
Upon suppression of the antiferromagnetic ZZ phase in 
a large
external magnetic field, the continuum acquires a gap
and the spectrum exhibits a sharp excitation that is smoothly connected to a spin-flip excitation in the field-polarized limit  \cite{sahasrabudhe2020high}. 
Altogether, the spectroscopic data reported thus far do not provide any signature of a distinct IFR
\cite{baek2017evidence, wang2017magnetic, ponomaryov2017unconventional,sahasrabudhe2020high, ponomaryov2020nature}. 
In Raman scattering at \SI{2}{\kelvin} in magnetic field, Wulferding \textit{et al}.\ \cite{wulferding2020magnon} 
observed an excitation which they interpreted as a Majorana bound state.  
However, the field was tilted by 18$^\circ$ out of the honeycomb plane and the measurements were restricted to \SI{10}{\tesla}.
Hence the data
could not capture any possible difference between the high-field and the intermediate field regime. 
Remarkably, this excitation has not been observed
in another Raman scattering study of \ru in high-magnetic fields \cite{sahasrabudhe2020high}, thus obstructing the clarity of opinion about the presence of an IFR.

Raman scattering is expected to be an effective probe for emergent Majorana fermions \cite{Nasu2016Raman, knolle2014Raman, Knolle2015structurefactor} and anyonic excitations. 
Experiments performed by Pinczuk \textit{et al}.\ have already demonstrated the strength of Raman scattering in probing the emergent anyonic excitations of the fractional quantum Hall effect \cite{Pinczuk1993magnetoroton,Kang2001anyonsbyRaman,Goni1993anyonsbyRaman}. 
Recently, it has been
suggested that these anyonic excitations are chiral and may show fingerprints in 
helicity-dependent Raman scattering  \cite{Ngyeun2021RamanQHE,Liou2019chiralgravitons}. 
Inspired by these ideas we performed helicity dependent Raman scattering on \ruu.
The blueprint for this is to compare the Raman intensities for circularly polarized light \cite{barron2004raman} 
in the four different polarization geometries
$I_{++}, I_{--}, I_{+-}, I_{-+}$, where the first (second) index denotes the circular polarization of the incident (scattered) light. As the continuum and the magnetic excitations in \ru do not scatter in $I_{++}$ and $I_{--}$ 
(see Supplementary Information \cite{SI}), we focus on the Raman Optical Activity (ROA) defined as
\begin{equation}
\text{ROA} = \frac{I_{+-}-I_{-+}}{I_{+-}+I_{-+}}\ .
\label{eq2}
\end{equation}
ROA is a detectable quantity that is intimately connected with the chirality of excitations \cite{barron2004raman,Chen2021ROA,Zhang2022ROA}. Chiral excitations possess handedness and are asymmetric under inversion. For example, let us consider magnons. Magnons in a ferromagnet are chiral and are right circularly polarized with respect to the direction of magnetization \cite{Jenni2022chirality}, while antiferromagnets have two degenerate magnon branches with opposite chirality. In the particular case of a ferromagnet, only one of the two incident 
polarizations $\sigma^{+}$ and $\sigma^{-}$ yields a Raman signal and this shows $\text{ROA}=1$
\cite{cenker2021direct,Hisatomi2019chiralmagnon,lyu2020probing}. In antiferromagnets,  an external magnetic field can lift the degeneracy of magnons which can again be detected using ROA \cite{Hoffman2005CircularDA}. 

In this letter, we focus on exploring the chirality of magnetic excitations in \ru through the associated ROA.\@
For an external magnetic field $B$ large enough to suppress the ZZ phase, we will show three different trends in ROA.\@
These will be pinpointed as characteristic fingerprints of three distinct field-induced magnetic regimes. 
One stunning example is the observation of a plateau of ROA\,=\,1 of the dominant magnetic excitation 
in the field range of \SIrange{7.5}{10.5}{\tesla} which serves as spectroscopic signature of
a prominent intermediate field regime.

\begin{figure}
\includegraphics[width=\linewidth]{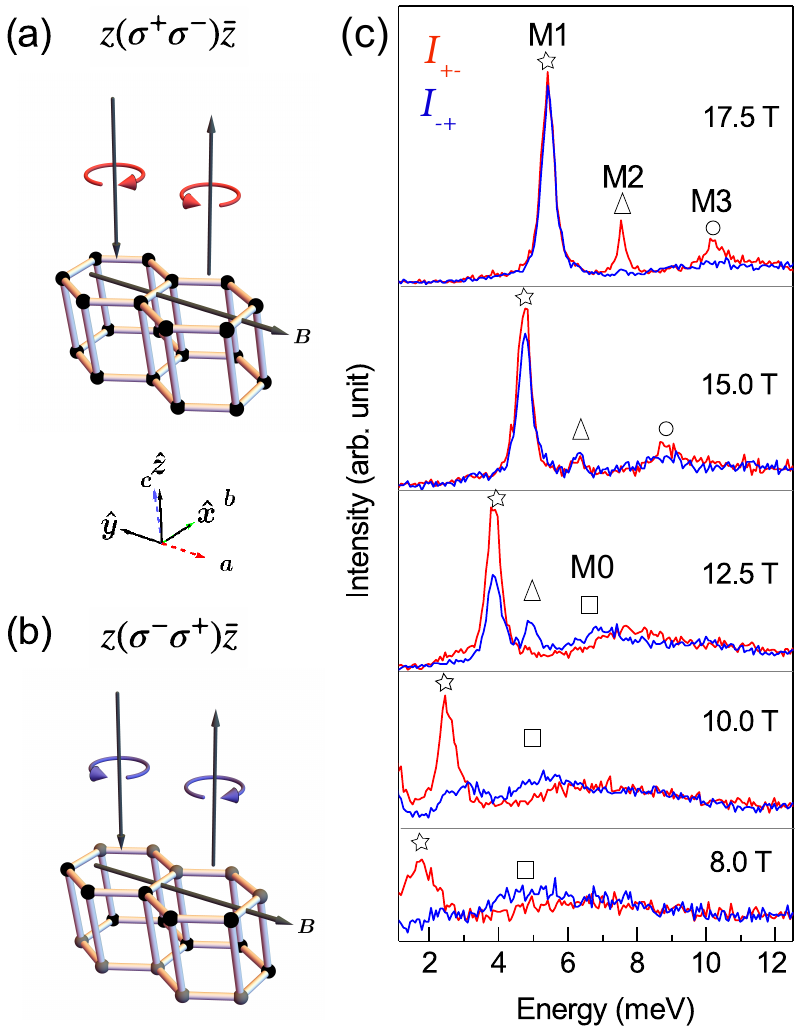}
\caption{Schematic of experimental geometry for measuring the helicity-resolved Raman intensities (a)  \LR  and (b)  \RL as plotted in (c)  for different magnetic field strengths. Square, star, triangle, and circle symbols denote the
magnetic excitations M0, M1, M2, and M3, respectively. }
\label{fig1} 
\end{figure}

%\textit{Methods:} 
For these experiments, high quality single crystals of \ru were prepared by the vacuum sublimation method \cite{SHDo2017majorana}.
The sample is placed in a magneto-cryostat and cooled down to \SI{1.7}{\kelvin}. The magnetic field 
is applied within the $ab$ plane, perpendicular to a Ru-Ru bond.
The light polarization is defined with respect to the $z$ axis. 
Left (right) circular light is represented as $\sigma^{+}$ $(\sigma^{-})$ as it projects spin angular momentum  
$+\hbar$ $(-\hbar)$ onto the quantisation axis. The experimental geometry is defined using the Porto notation
as $z(\sigma^{+}\sigma^{-})\bar{z}$ and $z(\sigma^{-}\sigma^{+})\bar{z}$, see Fig.~\ref{fig1}a, b.
We refer to the corresponding Raman intensities as $I_{+-}$ and $I_{-+}$, respectively. 
The temperature-dependent measurements have been performed by varying the power of the incident light so as to avoid a mechanical drift of the sample, see Supplementary Information \cite{SI}.

%\textit{Results:} 
\footnotetext[1]{The features M1, M2, and M3 correspond to $m_{1\alpha}$, $m_{2\gamma}$, and $m_{2\alpha}$ in Ref.\ \cite{sahasrabudhe2020high} and to M1, 2M, and 3M in Ref.\ \cite{wulferding2020magnon}, while M0 corresponds to MB in Ref.\ \cite{wulferding2020magnon}. Note that previous data were reported in \LR \cite{sahasrabudhe2020high} and \RL channels \cite{wulferding2020magnon}. Due to the pronounced ROA for intermediate fields, not all features are discussed in all previous studies.}

Figure\ \ref{fig1}c shows the helicity-dependent Raman response for $B>B_{\text{c}}$ (=\SI{7.5}{\tesla}). Overall, the observed features agree with previous Raman results \cite{sahasrabudhe2020high,wulferding2020magnon}. 
The data show the magnetic modes M0, M1, M2, and M3, and a broad magnetic multi-particle continuum. 
The M0 peak has been observed by Wulferding \textit{et al} \cite{wulferding2020magnon, Note1}

in a tilted magnetic field up to \SI{10}{\tesla} with incident $\sigma^{-}$ polarization.
They attributed this feature to a singlet Majorana bound state as its excitation energy remained insensitive 
to the magnetic field. 
The M1 mode has been identified as a spin-flip excitation with $|\Delta S|$\,=\,$1$ in the fully spin-polarized limit, i.e., for infinite magnetic field,
while M2 has been attributed to a two-particle bound state \cite{sahasrabudhe2020high,wulferding2020magnon}. The feature M3 has been interpreted either as a two-particle excitation since its energy at high fields roughly equals twice the energy of M1 \cite{sahasrabudhe2020high} or as a three-particle bound state \cite{wulferding2020magnon}. 
We emphasize that this  picture applies to the high-field limit  \cite{sahasrabudhe2020high,wulferding2020magnon,ponomaryov2020nature}. 

\begin{figure}[t]
	\includegraphics[width=\linewidth]{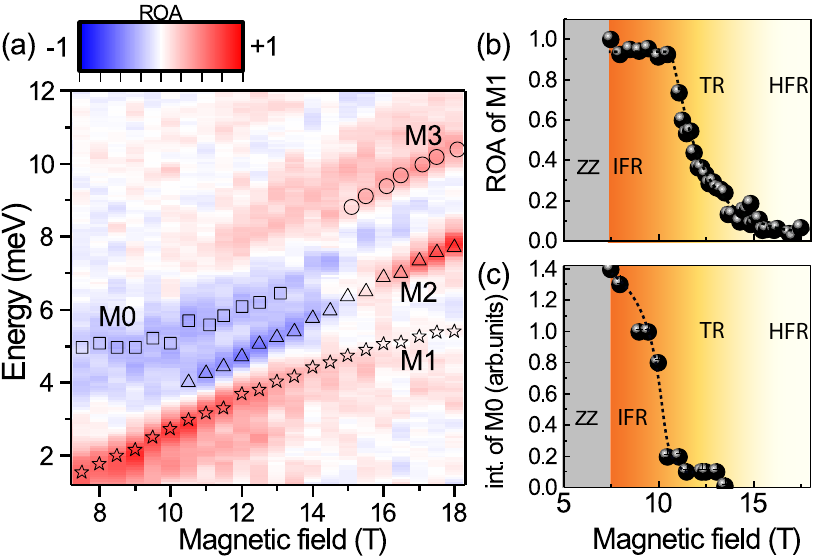}
	\caption{(a) 
ROA as a function of energy and field. 
Symbols denote the peak energies of M0, M1, M2, and M3.
 (b)  ROA of the M1 mode. 
The different trends in the ROA of M1 can be used 
to determine the three distinct magnetic-field regimes.
 Orange region: IFR with ROA\,=\,1; color gradient: transition regime (TR); white region: high-field regime (HFR) with ROA\,$\approx$\,$0$.  The grey shaded region indicates the ordered ZZ phase. Dashed lines: guide to the eye. (c)  Integrated intensity of the M0 mode as a function of magnetic field. }	\label{fig2}
\end{figure}

From the helicity dependence, it is immediately clear that the magnetic features M0, M1, M2, and M3 exhibit chirality. 
The effect of magnetic field on the chirality of these excitations 
allows us to distinguish different regimes. To this end, Fig.\ \ref{fig2}a depicts the ROA on a color scale.
We first focus on the ROA of M1, which can be continuously traced down 
to $B_{c}$. Values for ROA are obtained 
via Eq.\ (\ref{eq2}), using the integrated intensities obtained by fitting a Lorentzian to the M1 mode 
  in $I_{+-}$ and $I_{-+}$ (see Supplementary Information \cite{SI}). The result is depicted in 
  Fig.\ \ref{fig2}b, which reveals three distinct
trends in the ROA of M1. We find a clear plateau with ROA\,$\approx$\,$1$ from \SIrange{7.5}{10.5}{\tesla}
in the IFR,  a continuous decrease in ROA in the transition regime 
from \SIrange{10.5}{15}{\tesla}, and 
ROA\,$\approx$\,$0$ in the high-field regime 
above \SI{15}{\tesla}.

Supporting evidence for the thus identified onset of the high-field regime at \SI{15}{\tesla} emanates 
from the behavior of M0, M2, and M3, see Fig.\ \ref{fig2}a. The ROA of M2 crosses zero around \SI{15}{\tesla} 
and undergoes a drastic change from ROA\,=\,$-1$ at \SI{10.5}{\tesla} to ROA\,=\,$+1$ at \SI{18}{\tesla}. 
Concerning  M3, a clear peak is only identified above \SI{15}{\tesla}, while a consistent broad feature with ROA$>0$ 
is visible from $\SI{10.5}{\tesla}$ (see Fig.\ \ref{fig1}c, \ref{fig2}a). 
In contrast, M0 is only detected below \SI{15}{\tesla} where it shows a negative ROA (cf.\ Fig.\ \ref{fig1}c, \ref{fig2}c). 
As we do not observe any pronounced changes in ROA of these features from \SI{15}{\tesla} to \SI{30}{\tesla}, see Supplementary Information \cite{SI}, we identify $B \sim \SI{15}{\tesla}$ as the onset of the high-field regime.

Having identified the high-field regime, we now discuss the observations of the IFR.\@ 
Apart from a fully chiral 
M1, the only other chiral excitation that is observed in this regime is M0.
The energy (open squares in Fig.\ 2a) and the integrated intensity (Fig.\ 2c) of the M0 mode 
From 7.5 to \SI{10.5}{\tesla}, the
energy of M0 remains constant, 
while the intensity shows a pronounced decrease.
Above \SI{10.5}{\tesla}, M2 starts to peak in the energy window of 
M0, requiring two Lorentzians to account for the presence of both peaks.
Simultaneously, the excitation energy of M0 starts to increase linearly with $B$ with almost the same slope as M2. 
In summary, 
all relevant excitations (M0, M1 and M2) show distinct changes at $\SI{10.5}\tesla$, providing clear 
	spectroscopic evidence for the existence of an IFR.

\begin{figure}[h!]
\includegraphics[width=\linewidth]{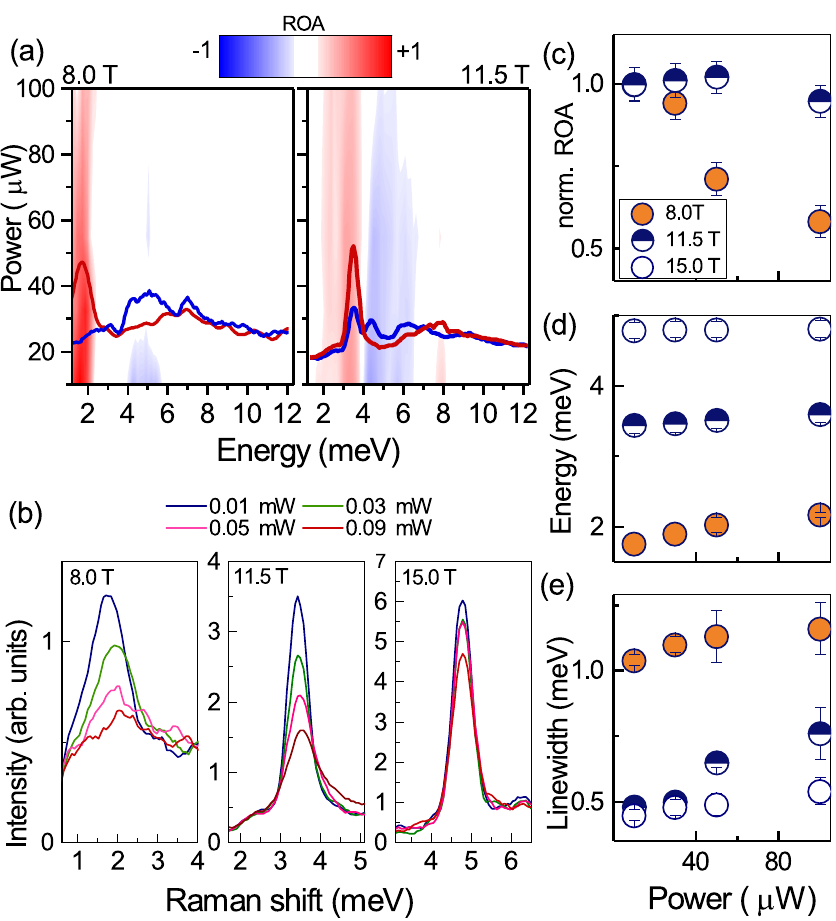}
\caption{(a) 
Power dependence of ROA
at \SI{8}{\tesla} and \SI{11.5}{\tesla}. Red (blue) areas represent larger scattering in $I_{+-}$ ($I_{-+}$). 
As a reference, the red and blue solid lines show 
\LR and \RL 
at \SI{10}{\micro\watt}. 
At \SI{8}{\tesla}, the M1 mode in the \LR channel shows maximum power sensitivity around \SI{40}{\micro\watt} ($T\sim\SI{7}{\kelvin}$), 
while
M0 in the \RL channel disappears around \SI{20}{\micro\watt} ($T <\SI{7}{\kelvin}$). 
(b) 
Effect of laser power on M1 in \LR in different field regimes. 
(c) Power-induced change in ROA of M1 
at \SI{8}{\tesla} and \SI{11.5}{\tesla}. 
(d),(e) Power dependence of energy and linewidth of the M1 mode in \LR channel. }

\label{fig3}
\end{figure}

Next we present our results from the measurements of the power-dependent spectra. Figure\ \ref{fig3}a exemplifies the power-dependent change of ROA at \SI{8}{\tesla} and \SI{11.5}{\tesla}, 
i.e., inside and just above the IFR, respectively. 
For ease of visualization, we have superimposed 
smoothened reference spectra 
taken
at \SI{10}{\micro\watt}.
At \SI{8}{\tesla}, the spectra show M1 and M0 around \SI{1.5}{\milli\eV} and \SI{5}{\milli\eV}, respectively, cf.\ Fig.\ 2a. At \SI{11.5}{\tesla}, the peaks are shifted to around \SI{3.2}{\milli\eV} and \SI{6.0}{\milli\eV} while M2 rears its head around \SI{4.5}{\milli\eV}. The color plot again pinpoints the opposite chiralities of M1 and M0.
The power dependence reveals a further key difference between \SI{8}{\tesla} and \SI{11.5}{\tesla}. In the IFR at \SI{8}{\tesla}, the 
chiralities of both M1 and M0 
rapidly
decline with increasing laser power, i.e., sample temperature. In contrast, the ROA of both modes is less sensitive to the laser power at \SI{11.5}{\tesla}. This is highlighted in Fig.\ \ref{fig3}c, showing the normalised ROA of M1. At \SI{40}{\micro\watt} ($\approx \SI{7}{\kelvin}$), the ROA of M1 is suppressed by more than a factor of two at \SI{8}{\tesla} but remains unaffected at \SI{11.5}{\tesla}. 
Figure\ \ref{fig3}b shows the smoothened spectrum of M1 as observed in $I_{+-}$, 
while Fig.\ \ref{fig3}d and \ref{fig3}e show the energy and the linewidth of M1 as a function of power for different fields, obtained by fitting M1 in $I_{+-}$  with a Lorentzian. %Again,
 The energy of M1 hardens with increasing laser power at \SI{8}{\tesla}.
Again, this behavior is found to be unique to the IFR.
Furthermore, the linewidth of M1 is much larger at \SI{8}{\tesla} than at higher fields.

%\textit{Discussion:} 
Previous Raman scattering studies on \ru in high magnetic fields reported on  $I_{+-}$ \cite{sahasrabudhe2020high} 
and 
 $I_{-+}$ \cite{wulferding2020magnon}. 
Due to the chiral character of the magnetic excitations, 
in particular with
M0 and M1 showing opposite chiralities, the key to a comprehensive picture of the magnetic Raman features and to a thorough understanding of the spectra lies in the comparison of the $I_{+-}$ and $I_{-+}$ channels, i.e., in the ROA.\@ 
In particular our ROA data measured over a wide range of magnetic fields above $B_\text{c}$ allow us to unravel the distinct spectroscopic characteristics of the intermediate-field regime and to elucidate its properties.

The IFR is 
characterized by the presence of the two
excitations M0 and M1 
for which we find opposite chiralities. Furthermore, ROA\,=\,$1$ for M1 within the IFR.\@
To illustrate the importance of M0 and M1 being chiral, we consider the case of a simple para\-magnet instead of a QSL above \SI{7.5}{\tesla}, where long-range zigzag order is suppressed by the magnetic field. The external magnetic field competes with exchange interactions. In this para\-magnet scenario, it is feasible to neglect field-induced order at $B_c$ and assume that the time-average of the local magnetic moments tends to zero. 
In this case, the Raman tensor $R$ of the excitations is defined by the crystallographic space group
$C2/m$ \cite{TTMai2019continuum} or $R\bar{3}$ \cite{Glazmada2017structure}. The Raman intensity is given by $I_{is}=|e_{s}^{\dag}\cdot R \cdot e_{i}|^{2}$, where $e_{s}$ and $e_{i}$  are the electric field vectors of the scattered and the incident light. 
Since neither of these crystallographic space groups is 
chiral, we find $I_{+-}$\,=\,$I_{-+}$ and ROA\,=\,0, see Supplementary Information \cite{SI}. 
The pronounced ROA of the magnetic excitations M0 and M1 in the IFR indicates an unconventional ground state distinct from a conventional paramagnet.

Previously, M0 has been attributed to a singlet Majorana bound state \cite{wulferding2020magnon}. 
While our results do not provide any direct evidence for a Majorana character, they support the interpretation in terms of a bound state to some extent.
Above \SI{10.5}{\tesla}, i.e., above the IFR, the M0 mode hardens and exhibits a strongly suppressed intensity,
hinting at an unbinding process, i.e., a decay into the multi-particle continuum. 
However, the large width of M0 speaks against a strict bound-state scenario but rather suggests a resonance within the continuum close to its lower edge.

The mode M1 has been identified as a spin-flip excitation in the high-field regime \cite{sahasrabudhe2020high}. 
Above \SI{15}{\tesla}, the ROA of M1 
is negligible.
The pronounced increase of the ROA with decreasing field and in particular the plateau with ROA\,=\,$1$ in the IFR are clear signatures of a change of character of this mode. 
Remarkably, the plateau is observed between \SIrange{7.5}{10.5}{\tesla}, consistent with the intermediate field regime where a QSL phase has been proposed to exist \cite{Czajka2021oscTQHE, yokoi2021half, kasahara2018majorana, yamashita2020sample, balz2019finite}. 
 The linewidth $\gamma$ of the M1 mode equals \SI{0.5}{\milli\eV} in the high-field regime where it corresponds to a spin flip, see Fig.\ \ref{fig3}e. In the IFR, $\gamma$ is enhanced by more than a factor of two. It is tempting to explain this increase of the linewidth in a scenario based on Majorana fermions, where the local spins decay into Majoranas and fluxes, giving rise to a substantial broadening of M1.
 
 Lastly, we briefly 
address
our observation of chiral multi-magnon excitations M2 and M3. Topological magnons contributing to the thermal Hall effect are expected to live in the field-polarized phase of a Kitaev-Heisenberg magnet \cite{Joshi2018topmag,McClarty2018topmag,Chern2021topmagnons}. Even though a detailed \ analysis of the chiral origin of these excitations is beyond the scope of this work, the chirality of these multi-particle excitations hints at their topological nature and indicates an onset of field polarization.

In conclusion, magnetic field and laser power dependent Raman 
optical activity experiments show the presence of distinct field induced 
phases for \ruu. 
In particular, the upper limit of our identified intermediate field regime is very close to the value of \SI{11}{\tesla}, above which the quantum oscillations intrinsic to a QSL phase vanish \cite{Czajka2021oscTQHE}.
% reported from the quantum oscillations  of the QSL phase \cite{Czajka2021oscTQHE}. 
The intermediate field regime
shows intriguing spectroscopic features.
The most salient 
features are a fully chiral response (ROA=1) of the strongest (M1) 
magnetic mode, a strong broadening of this mode 
may suggest a fractional character of the fundamental magnetic excitations 
specific for the IFR,
and the presence of a resonance mode (M0) on the lower 
edge of the magnetic Raman continuum which previously has been discussed 
in terms of a Majorana bound state \cite{wulferding2020magnon}. The laser power 
dependence (i.e., temperature dependence) of the spectra indicates that 
the ground state in the IFR region is rather fragile as is demonstrated 
by the rapid decrease of the fully chiral response upon increasing 
temperature. 
In contrast,
the ROA of the spectra in the high field regime 
is hardly affected by the increased temperature. 
Though it is clear from the present and earlier experiments that the IFR 
has an unconventional ground state which could very well be a quantum 
spin-liquid state, its exact nature remains elusive at this point.
The present work shows the power 
of ROA experiments in materials showing unconventional magnetism and 
sheds light on the intriguing chiral magnetic properties of the 
Kitaev-Heisenberg magnet \ruu, inspiring future research 
to realize a field-induced spin-liquid phase in \ru and 
other Kitaev-like systems.
\newline

%\section*{Acknowledgments}
\begin{acknowledgments}
We thank Aprem Joy, Ciar\'an Hickey, David Kaib, Roser Valentí,  Daniel I. Khomskii and Fulvio Parmigiani  for insightful discussions. 
We thank Jonathan Buhot and Cl\'ement Faugeras for supporting preliminary experimental measurements.
A~S thanks Vivek Lohani for critical comments on the manuscript. M~P thanks Matija \v{C}ulo and Peter C. M. Christianen for support during the experiment. 
The authors acknowledge financial support by the Deutsche Forschungsgemeinschaft
(DFG, German Research Foundation) via Project No.\ 277146847 -- CRC 1238 
(project B03), via Project
No.\ 107745057 -- TRR 180 
(project F5), 
and via Project No.\ VA117/15-1. 
A~L and V~T acknowledge support by the 
DFG through TRR 80.
V~T acknowledges the support via the project ANCD 20.80009.5007.19 (Moldova).
 \end{acknowledgments}

\def\bibsection{\section*{~\refname}} 

%\bibliography{bibliography}

%\bibliography{bibliography_main}


\end{document}